\documentclass[aps,reprint,prr,longbibliography,a4paper,twocolumn,superscriptaddress,citeautoscript]{revtex4-1}
\usepackage[normalem]{ulem}
\usepackage[svgnames]{xcolor}
\usepackage[colorlinks=True,linkcolor=DarkRed,citecolor=ForestGreen,urlcolor=MediumBlue]{hyperref}
\usepackage[english]{babel}
\usepackage{lmodern}
\usepackage[T1]{fontenc}
\usepackage{amsmath}
\usepackage{units}
\usepackage[pdftex]{graphicx}
\usepackage{grffile}
\usepackage{color}
\usepackage[bottom]{footmisc}
\usepackage{graphicx}
\usepackage{dcolumn}
\usepackage{bm}\let\vec\bm 
\usepackage{soul}
\usepackage{amsfonts}
 \usepackage{braket}
\usepackage{amssymb}
\usepackage[figuresright]{rotating}
\usepackage{amsfonts}
 \usepackage{braket}
 \usepackage[normalem]{ulem}
\usepackage{hyperref}
\usepackage{color}
\usepackage[english]{babel}
\usepackage{lmodern}
\usepackage[T1]{fontenc}
\usepackage{amsmath}
\usepackage{units}
\usepackage{grffile}
\usepackage{color}
\usepackage[bottom]{footmisc}
\usepackage{graphicx}
\usepackage{dcolumn}
\usepackage{bm}
\usepackage{soul}
\def\Xint#1{\mathchoice
   {\XXint\displaystyle\textstyle{#1}}
   {\XXint\textstyle\scriptstyle{#1}}
   {\XXint\scriptstyle\scriptscriptstyle{#1}}
   {\XXint\scriptscriptstyle\scriptscriptstyle{#1}}
   \!\int}
\def\XXint#1#2#3{{\setbox0=\hbox{$#1{#2#3}{\int}$}
     \vcenter{\hbox{$#2#3$}}\kern-.5\wd0}}

\def\dashint{\Xint-}
\begin{document}
\title{Thoughts about boosting superconductivity}
\author{Dirk van der Marel}

\address{Department of Quantum Mater Physics, University of Geneva, 24 Quai Ernest-Ansermet, 1211 Geneva 4, Switzerland
\\
\vspace{1\baselineskip}
This manuscript version is made available under the CC-BY-4.0 license 
https://creativecommons.org/licenses/by/4.0/
\vspace{1\baselineskip}
}

\date{February 7, 2025}
\begin{abstract}
In a superconductor electrons form pairs despite the Coulomb repulsion as a result of an effective attractive interaction mediated by, for example phonons. In the present paper DeGennes' description of the dynamically screened Coulomb interaction is adopted for the effective interaction. This model is generalized by including the elastic response of the charge-compensating background and the BCS gap equation is solved for the resulting effective electron-electron interaction. It is demonstrated that the superconducting critical temperature becomes strongly enhanced when the material is tuned close to a structural instability.  
\end{abstract}

\maketitle

\section{Introduction}
\label{section:introduction}
Jan Zaanen took no interest in boring subjects. Our first conversation in the autumn of 1982 included superconductivity, the Kondo effect and general relativity. Jan and I had just started our Ph D studies with George Sawatzky and we were making mutual introductions in our shared office in the basement of the chemistry building at the University of Groningen. This conversation was the first one of countless scientific discussions and it kicked off a lasting friendship.

One of the major questions that has occupied the generation to which Jan and I belonged, was inspired by the discovery of high temperature superconductivity in the cuprates~\cite{bednorz1986}. 
In BCS theory a pairing instability occurs at a critical temperature $T_c$ where $\lambda{\chi}^{\prime}_{0}(T_c)=1$. Here $\lambda$ is an attractive pairing interaction, ${\chi}^{\prime}_{{0}}(T)\sim\ln(\omega_{{0}}/k_BT)$ is the static bare pair susceptibility and $\omega_{{0}}$ is an appropriately weighted average energy of the phonons involved in the pairing interaction. The conventional approach has been to argue that in the cuprates bosonic degrees of freedom such as phonons, plasmons, spin-fluctuations, loop-currents or excitons mediate a strong attractive pairing interaction.
In a series of papers with Jian-Huang She and others, Jan and his coauthors argued~\cite{she2009,she2011,yang2011,she2013}, that the bare pair susceptibility in the cuprates has properties qualitatively different from standard metals, causing $T_c$ to be very high even when the pairing interaction isn't stronger than in other superconducting materials. They started from the Ansatz that the bare pair susceptibility of the cuprates has an algebraic temperature dependence ${\chi}^{\prime}_{{0}}(T) \sim 1/T^{\alpha}$ and this, they reasoned, causes the pairing instability to occur at relatively high temperatures. This Ansatz was motivated by the results of Ref.~\cite{marel2003} on the “Planckian” nature of the relaxation time observed in optical experiments resulting from an animated discussion on our way to a meeting in Poland. 

The concept that, as a result of coupling to phonons~\cite{bardeen1955} or other collective degrees of freedom such as spin fluctuations~\cite{scalapino1986} or loop currents~\cite{varma2014}, electrons overcome the Coulomb barrier and form pairs, is inherently counter intuitive. The -to my taste- most intuitively appealing explanation of how this works, was given by DeGennes in his textbook on superconductivity~\cite{degennes1966} and was based on the “jellium model”. It turned out that solving this model is not entirely trivial~\cite{ginzburg1968,kirzhnits1969,marel-berthod2024} and a number of questions remain, some of which I will discuss below.
\section{The jellium model}
\label{section:jellium}
\begin{figure}[]
\includegraphics[width=\columnwidth]{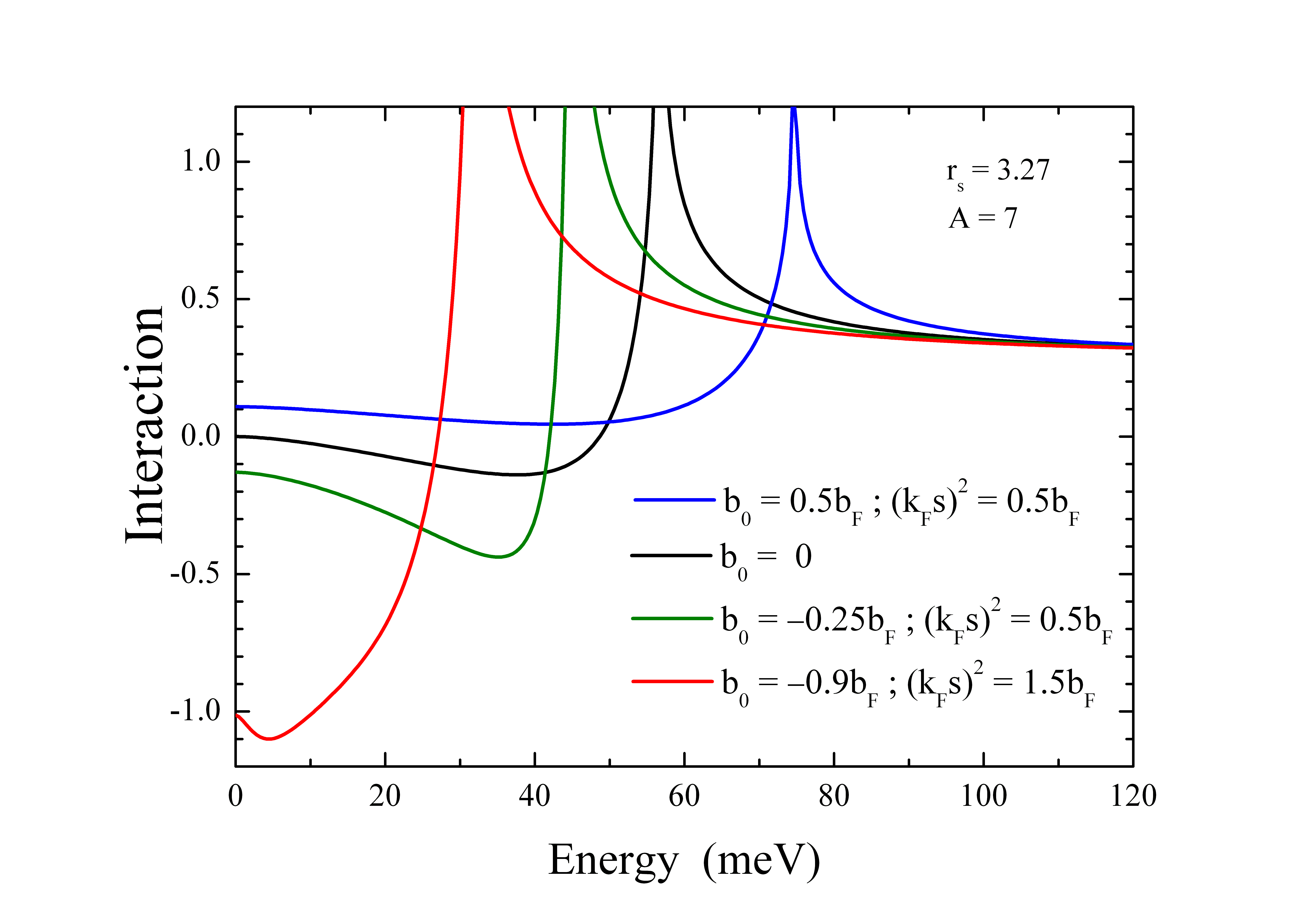}
\caption{\label{fig:upsilon} Interaction potential in the s-wave channel, Eq.~\ref{eq:upsilon_s}, as a function of $\hbar\omega$ in the case of lithium for four representative values of the reduced bulkmodulus of the charge-compensating background, $b_0$, and of $(k_Fs)^2$. For this $r_s$ value the reduced bulk modulus of the electrons is $b_F=\pi k_F a_0/4 = 0.461$. The curves displayed here are for $k=k^{\prime}=k_F$. The Fermi energy is $\epsilon_F=4.7$~eV and the plasma energy of the charge-compensating background is $\hbar\omega_{{0}}=70$~meV. }
\end{figure}

The interaction process where two electrons are scattered from $\ket{\vec{k},-\vec{k}}$ to $\ket{\vec{k}^{\prime},-\vec{k}^{\prime}}$ can be described by considering the Coulomb interaction $V(q)$ and the screening thereof~\cite{degennes1966}:
\begin{align}
V^{\mathrm{eff}}(q,\omega) 
= \frac{V(q)}{\epsilon(q,\omega)},
\label{eq:veff}
\end{align}
where $q=|\vec{k}-\vec{k}^{\prime}|$. In the jellium model the potential landscape of the charge-compensating background is flat, so that the energy-momentum relation is $\epsilon=\hbar^2k^2/2m_e$. The dielectric function of a system of free electrons in a compressible positively charged background is (see~\ref{section:response} )
\begin{align}
\epsilon(q,\omega)&=1+\frac{k_{0}^2}{q^2}+
\frac{\omega_{{0}}^2}{B_0\rho_m^{-1} q^2/(1+q^2s^2)-\omega^2}
\label{eq:epsilon}
\end{align}
where $k_0=\sqrt{4k_F/(\pi a_{{0}})}$ is the Thomas-Fermi wave vector, $\omega_{{0}}$ is the plasma frequency of the charge-compensating background, $b_0$ is it's bulk modulus, $\rho_m$ it's mass density and $s$ is the length scale below which $B_q$ vanishes. 
For our discussion it is useful to define {\it reduced} parameters for the bulk moduli of the charge-compensating background ($b_0$) and of the electrons ($b_F$) 
\begin{align}
b_0 &= B_0\frac{k_F^2}{\rho_m\omega_{{0}}^2}
\label{eq:b0}
\\
b_F &= \frac{k_F^2}{k_{0}^2} = \frac{ \pi k_Fa_{{0}}}{4}
\label{eq:bF}
,
\end{align}
both of which are dimensionless constants.
\onecolumngrid
To obtain the effective interaction in dimensionless form we multiply $V^{\mathrm{eff}}(q,\omega)$ with the density of states at the Fermi level
\begin{align}
N(0) &= \frac{m_e k_F}{2 \pi^2 \hbar^2}.
\end{align}
The effective interaction in the s-wave channel is then given by
\begin{align}
\upsilon^{s}_{k,k^{\prime}}(\omega)=\frac{1}{2} \int_{{0}}^{\pi}N(0) V^{\mathrm{eff}}(q,\omega) \sin{\theta}d\theta,
\label{eq:definition_upsilon_s}
\end{align}
where $\theta$ is the angle between $\vec{k}$ and $\vec{k^{\prime}}$. The result  of this integral in closed form is 
\begin{align}
\upsilon^{s}_{k,k^{\prime}}(\omega) 
&=
\upsilon^{s,1}_{k,k^{\prime}}(\omega) 
+
\upsilon^{s,2}_{k,k^{\prime}}(\omega) 
&
\label{eq:upsilon_s}
\\
\upsilon^{s,1}_{k,k^{\prime}}(\omega) &=
\frac{1}{8\kappa}
\left[\sqrt{g}-\frac{h}{\sqrt{g}}\right]
\ln\left|
\frac{({e}+\kappa_{+})({e}+\kappa_{-})-{{g}}-2{\kappa} {\sqrt{{g}}}}{({e}+\kappa_{+})({e}+\kappa_{-})-{{g}}+2{\kappa} {\sqrt{{g}}}}
\right| &(g>0)
\\
\upsilon^{s,1}_{k,k^{\prime}}(\omega) &=
\frac{1}{4\kappa}
\left[\sqrt{-g}+\frac{h}{\sqrt{-g}}\right]
\arctan\left[\frac{2\kappa\sqrt{-g}}{(e+\kappa_{+})(e+\kappa_{-})-g}\right] 
&(g<0)
\\
\upsilon^{s,2}_{k,k^{\prime}}(\omega) &=
\frac{1}{16\beta_F{\kappa} }
\ln\left|\frac{({e}+\kappa_{+})^2-{{g}}}{({e}+\kappa_{-})^2-{{g}}}\right|,
&
\end{align}
where
\begin{align}
{\kappa}&=\frac{kk^{\prime}}{k_F^2 } 
;\hspace{2mm}
{\kappa_{\pm}}=\frac{(k\pm k^{\prime})^2}{2k_F^2} 
;\hspace{2mm}
{e}=e_1+e_2
;\hspace{2mm}
e_1=\frac{1-\beta_F z^2/\zeta_0^2}{4\beta_0}
;\hspace{2mm}
e_2=\frac{1}{4\beta_F}
;\nonumber\\
{g}&=e^2+\frac{z^2/\zeta_0^2}{4\beta_0}
;\hspace{2mm}
{h}=e_1^2-e_2^2
;\hspace{2mm}
\zeta_0^2 = \frac{b_0 z_0^2}{b_0 -  b_F k_F^2s^2 z^2/z_0^2} 
;\hspace{2mm}
z_0^2=\frac{4m_e}{3m_N}
;\hspace{2mm}
z = \frac{\hbar \omega }{\epsilon_F}
;\nonumber\\
\beta_F&=b_F\frac{b_0 +k_F^2s^2(1-b_Fz^2/z_0^2)}{b_0 -   b_F k_F^2s^2z^2/z_0^2} 
;\hspace{2mm}
\beta_0=[ b_0 -  b_F k_F^2s^2  z^2/z_0^2]\times[1+k_F^2s^2b_0^{-1}(1-b_Fz^2/z_0^2)]
.
\label{eq:factors}
\end{align}
This function is displayed in Fig.~\ref{fig:upsilon} for mass number $7$, density $n=4.6\cdot 10^{22}$~cm$^{-3}$ and a selection of values of $b_{0}$ and $s$. 
For $\omega\rightarrow\infty$ the interaction converges to the Thomas-Fermi screened Coulomb repulsion.
The interaction has an explicit dependence on momentum and frequency, which is handled in different ways depending on the formalism:
\begin{enumerate}
\item\label{item:BCS} If one solves the BCS gap equation~\cite{bardeen1957} with the Bardeen-Pines approach~\cite{bardeen1955}, the $\omega$ dependence is treated as the on-shell interaction, $\hbar\omega=\epsilon_{\vec{k}}-\epsilon_{\vec{k}^{\prime}}$. 
\item If one solves the Eliashberg equations~\cite{eliashberg1960}, the dependence on $\omega$ is transformed to Matsubara frequencies which are on the imaginary frequency axis. The singularities seen in Fig.~\ref{fig:upsilon} are absent for the imaginary axis, which has certain practical advantages for numerical coding. After solving the gap-function for the Matsubara frequencies one can in principle reconstruct the gap-function along the real axis, which is however a challenging numerical procedure. 
\item If one solves the McMillan equation~\cite{mcmillan1968}, one usually starts from the electron-phonon coupling function $\alpha^2F(\omega)$ which is nowadays typically obtained from a frozen-phonon band structure calculation. Formally the electron-phonon coupling function and the pairing interaction are connected by Kramers-Kronig relations, {\it i.e.} 
$\upsilon(\omega)=\mu-2\dashint_0^{\infty}\frac{\nu\alpha^2F(\nu)}{\nu^2-\omega^2}d\nu$,~\cite{marsiglio2020} where $\mu$ is the the Thomas-Fermi screened Coulomb repulsion. In the MacMillan approach the phonon-mediated interaction $\lambda$ is calculated from integrating $\alpha^2F(\omega)/\omega$. 
\end{enumerate}
In section~\ref{section:superconductivity} we will solve the BCS gap equation, {\it i.e.} item~\ref{item:BCS} of the list above. Before doing so we open a parenthesis to briefly discuss a frequently used approximation which we will {\it not} use in section~\ref{section:superconductivity}. This approximation rests on the argument that the most important contribution to the pairing interaction originates from the Fermi surface area. The Coulomb interaction can then be represented by a single parameter $\mu$, corresponding to $k=k^{\prime}=k_F$
\begin{align}
\mu  &= \frac{1}{8 b_F }
\ln\left(1+4 b_F \right).
\label{eq:mu}
\end{align}
Continuing the discussion of the case $k=k^{\prime}=k_F$, the static ($\omega = 0$) interaction is
\begin{align}
\upsilon^{s}(0)  &=  
\frac{1}{8 b_F }\frac{b_0}{k_F^2s^2+b_0}
\ln\left|1+
\frac
{4 b_F(k_F^2s^2+b_0) }
{b_F+b_0}
\right|,
\end{align}
which combines the phonon-mediated attraction and the Coulomb repulsion. Correspondingly, the parameter characterizing the phonon-mediated attraction is 
\begin{align}
\lambda=\mu-\upsilon^{s}(0)
.
\end{align}
\twocolumngrid
We see right away that for $b_0=0$ we have $\lambda=\mu$, for $b_0>0$ we obtain $\lambda<\mu$ and for $b_0<0$ it follows that $\lambda>\mu$. Fig.~\ref{fig:upsilon} illustrates these three cases. At this point we close the parenthesis about the $k=k^{\prime}=k_F$ approximation. In section~\ref{section:superconductivity} we solve the BCS gap equation with the the substitution $\hbar\omega=\epsilon_{\vec{k}}-\epsilon_{\vec{k}^{\prime}}$ and the full $k$ and $k^{\prime}$ dependence of the interaction. 
\begin{figure}[t]
\includegraphics[width=\columnwidth]{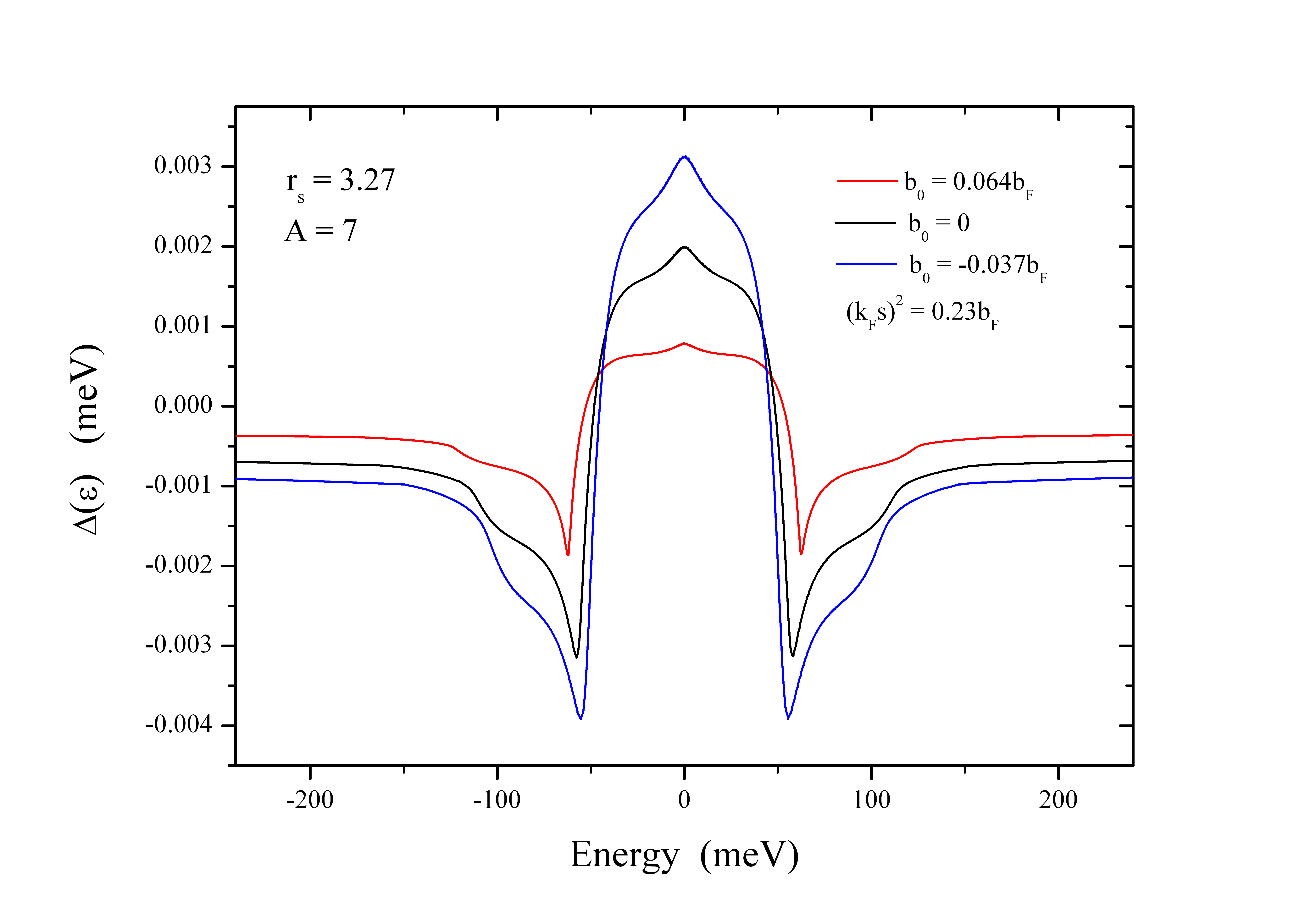}
\caption{\label{fig:gap_function} Gapfunction at $T=0$ in the case of lithium for three different values of the reduced bulk modulis $b_0$ and $(k_F s)^2=0.23 b_F$. For this $r_s$ value the reduced bulk modulus of the electrons is $b_F= 0.461$. }
\end{figure}
\section{Stability considerations}
\label{section:stability}
The sound dispersion of a metal is obtained from the condition $\epsilon(q,\omega)=0$. Taking the limit $q\rightarrow 0$, we obtain the sound velocity:
\begin{align}
c=\frac{\omega}{q} =\frac{\omega_{{0}} }{k_F}\sqrt{b_F+b_0}.
\end{align}
Correspondingly, the overall reduced bulk modulus of the charge-compensating background and the electrons together is
\begin{align}
b_{tot} =  b_F+b_0.
\end{align}
For $b_0>-b_F$ the overall compressibility is positive. 
While this is a necessary condition for stability of the system, it is not a sufficient one. A stricter requirement is, that $\mbox{Re}\epsilon(q,0)$ should {\it not} be in the range $\{0;1\}$~\cite{dolgov1981}. Consequently, from the expression for the dielectric function, Eq.~\ref{eq:epsilon}, for the present model the stability condition is
\begin{align}
b_0>-\min{(k_F^2s^2,b_F)}
\end{align}
We furthermore notice that $1/\epsilon(q,0)$ diverges for $q=q_{c}$ where
\begin{align}
q_{c}= k_F \sqrt{\frac{- b_F-b_0}{k_F^2 s^2 + b_0} }.
\end{align}
The square root is real for $b_0\in\{-b_F ;-k_F^2s^2\}$, implying that in this parameter range the system is unstable with respect to a charge density wave (CDW) of wave vector $q_{c}$. To evaluate the superconducting properties in this range requires knowledge about the amplitude of the CDW in equilibrium and of its impact on the electron-dispersion, which is outside the limitations of the present model.
\begin{figure}[t]
\includegraphics[width=\columnwidth]{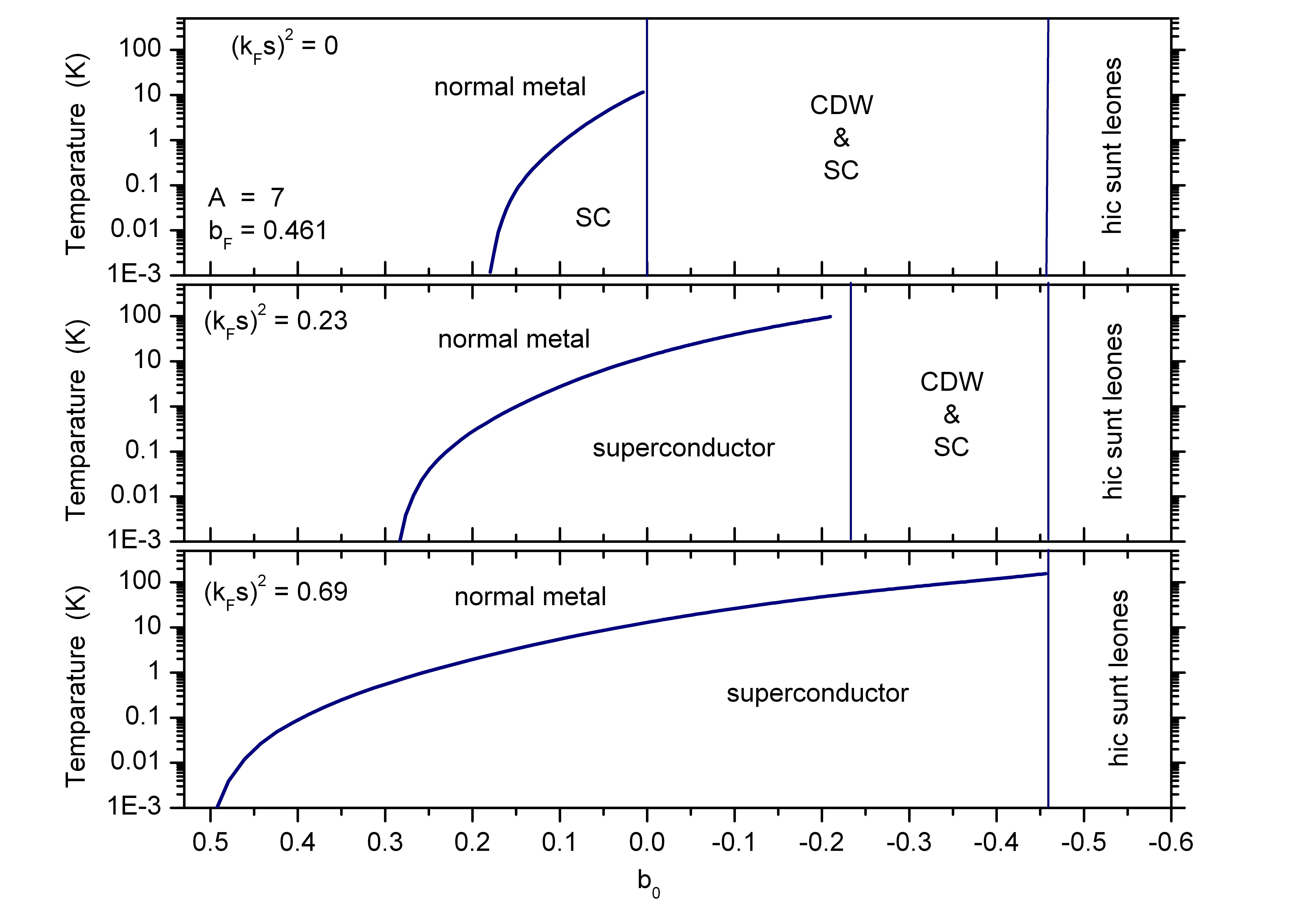}
\caption{\label{fig:tc_b0} Phase diagram in the case of lithium for 3 different values of $(k_Fs)^2$. The horizontal axis is the reduced bulk modulus of the charge-compensating background, Eq.\ref{eq:b0}. Middle panel ($k_F^2 s^2=0.23$): In principle the curve extends to $b_0=-0.23$, but for $-b_0\lessapprox 0.23$ the numerical output of the gap equation becomes inaccurate.}
\end{figure}
\section{Superconductivity}
\label{section:superconductivity}
The BCS gap equation for s-wave pairing, using the corresponding definition of the interaction, Eq.~\ref{eq:definition_upsilon_s}, is
\begin{align}
&\Delta({\epsilon})=-
\int_{{0}}^{\infty} 
\frac{\upsilon^{s}_{k,k^{\prime}}(\omega)\Delta({\epsilon^{\prime}})}{2\sqrt{(\epsilon^{\prime}-\mu_F)^2+|\Delta({\epsilon^{\prime}})|^2}}\times \nonumber\\&
\times\tanh\left(\frac{\sqrt{(\epsilon^{\prime}-\mu_F)^2+|\Delta({\epsilon^{\prime}})|^2}}{2k_B T}\right)
\sqrt{\frac{\epsilon^{\prime}}{\epsilon_F}}
d\epsilon^{\prime}
,
\label{eq:BCS}
\end{align}
where $\hbar\omega=\epsilon-\epsilon^{\prime}$ and $\mu_F$ is the chemical potential. The density of states is $N(\epsilon)=N(0)\sqrt{\epsilon/\epsilon_F}$, but note that the factor $N(0)$ has been absorbed in the definition of the interaction, Eq.~\ref{eq:definition_upsilon_s}. 
Since in the present study $k_BT_c \ll \epsilon_F$, it follows that $\mu_F\cong \epsilon_F$. A consequence of Eq.~\ref{eq:BCS} is, that $\Delta({\epsilon})$ can be chosen real. For display purposes we adopt the convention that  $\Delta(0)$ is real and positive.
As illustrated in Fig.~\ref{fig:upsilon}, if ${b_0}= 0$, the effective interaction has a negative sign for energies well below $\hbar\omega_{{0}}$. On the other hand, $\upsilon^{s}(0)  = 0$, {\it i.e.} the {\it static} interaction is zero. Given this state of affairs one may wonder if the ground state is superconducting. This question was addressed for the case of hydrogen, also using the jellium model, by Ginzburg and Kirzhnits~\cite{ginzburg1968}, by Kirzhnits~\cite{kirzhnits1969} and in a recent paper of the present author and Berthod~\cite{marel-berthod2024}. The answer was that, in fact, the ground state is superconducting. While the original estimates were in the range of a few 100~K~\cite{ginzburg1968,kirzhnits1969}, we found relatively low values of $T_c$ with this model~\cite{marel-berthod2024}: The density dependence of $T_c$ is dome shaped with the maximum, $T_c=30$~K, at $r_s=3.5$. Although this is far below the original estimates~\cite{ginzburg1968,kirzhnits1969}, still the message of principle remains that, despite the fact that $\upsilon^{s}(0)= 0$, there is a non-trivial superconducting solution. 
Cohen and Anderson~\cite{cohen1972} argued that the static dielectric function cannot be negative. Consequently $\upsilon^{s}(0) \ge 0$, which is equivalent to the condition $\lambda\le \mu$. They simplified the interaction potential to $V(k,k^{\prime})=\lambda-\mu$ for energies below $\omega_0$, $V(k,k^{\prime})=-\mu$ for energies above $\omega_0$ and obtained
\begin{align}
k_B T_c& = \frac{2 e^{\gamma}}{\pi}\omega_{0} \exp{\left(-\frac{1}{\lambda-\mu^*}\right)}
\label{eq:cohen_anderson}
\end{align}
where $\mu^*=\mu/[1+\mu\ln(\epsilon_F/\omega_{0})]$ is the screened Coulomb pseudopotential.
With the aforementioned constraint that the static dielectric function cannot be negative, the additional constraint that $\mu<1/2$, and taking realistic values for the Fermi energy and the phonon plasma frequency ($\epsilon_F\sim 7$~eV, $\omega_0\sim 0.1$~eV) they concluded that $T_c$ has a maximum of about $10$~K. However, as discussed in the previous section, Dolgov, Khirzhnits and Maksimov~\cite{dolgov1981} have demonstrated that negative values of the static dielectric function are possible, but values in the range $\{0;1\}$ cannot occur for a stable system.

\vspace{1\baselineskip}
We now turn to the case of lithium. The cell volume of solid lithium is $V_{{0}}=21.6$~\AA$^3$~\cite{hanfland1999}. Consequently the conduction-electron density is $n=4.6\cdot 10^{22}$~cm$^{-3}$ and the Wigner-Seitz radius is $r_s=3.27$. Correspondingly $k_F =1.1  \cdot 10^8$~cm$^{-1}$, $\epsilon_F=4.7$~eV and $\hbar\omega_{{0}}=70$~meV. In Fig.~\ref{fig:gap_function} the gap function is shown for 3 representative cases corresponding to the interaction potential displayed in Fig.~\ref{fig:upsilon}, $b_0$ positive, zero and negative. We see that the effect of positive $b_0$ is to suppress the superconducting order. In contrast, a negative $b_0$ enhances $T_c$. Regarding positive $b_0$ it is of interest to point out that in this case the interaction (blue curve in Fig.~\ref{fig:upsilon}) is entirely repulsive for all energies. Indeed, as illustrated by Eq.~\ref{eq:cohen_anderson}, even in this case a superconducting solution can be obtained, provided that $\lambda>\mu^*$~\cite{coleman2015}. In this case the gap $\Delta(\omega)$ changes sign exactly where the Coulomb interaction becomes repulsive as illustrated in Fig.~2 of Ref~\cite{morel1962} and Fig.~\ref{fig:gap_function} of the present paper.
 
In Ref.~\cite{marel-berthod2024}, it was shown that $T_c^* = 0.925 \sqrt{\varepsilon_{\mathrm{F}}E_{c}} /k_{\mathrm{B}}$ where $E_c$ is the condensation energy calculated at $T=0$, coincides with the critical temperature following from the jump in the specific heat, providing an efficient method for obtaining the critical temperature. 
In Fig.~\ref{fig:tc_b0} the critical temperature calculated from the condensation energy is displayed as a function of $b_0$. 
Pointing to the left is the direction of positive $b_0$ where the critical temperature decreases below $1$~mK. Experimentally, the critical temperature has been determined as $T_c=0.4$~mK~\cite{tuoriniemi2007}. In view of these findings it seems plausible that the low $T_c$ values in the alkali metals are caused by none-negligible elasticity of the charge-compensating background.
Pointing to the right is the direction of negative $b_0$, illustrating that tuning the system toward an elastic instability is a promising strategy for boosting $T_c$. 
Note that the acoustic phonons soften in this limit. Using theoretical models different from the one employed here, Bergmann and Rainer~\cite{bergmann1973}, Maksimov and Savrasov~\cite{maksimov2001} and Jiang {\it et al.}~\cite{jiang2023} have also concluded that phonon softening should typically be accompanied by an increase of $T_c$.

While we have seen that a negative $b_0$ is beneficial for superconductivity, the elastic response in this model is given by $b_0+b_F$. If a negative value of $b_0+b_F$ were to occur, the material would collapse. For this reason candidates for this type of boosting of superconductivity have to be materials in proximity to a lattice instability, {\it i.e.} with $b_0<0$ while still being in a thermodynamically stable state with $b_0+b_F>0$. It is tempting to look for a relation to the observed $T_c$ enhancement of lithium up to 20~K for pressures in the range of 20 to 70 GPa~\cite{shimizu2002,deemyad2003}. However, at high densities the overlap of core electrons becomes an important factor and contrary to intuition the electronic structure of lithium becomes less free-electron like~\cite{neaton1999}, atoms form pairs and above 80 GPa the material is even semiconducting~\cite{matsuoka2018}. The high pressure phases of lithium are therefore not captured by the model here considered. That said, in a realistic description of the system taking into account the lattice structure, instabilities can occur similar in nature as the one that we discussed above. For the same reason the dielectric function could become negative when the system is tuned close to such an instability, and once again the static interaction would be attractive with the effect of boosting $T_c$. In that sense, the structural instabilities that are known to occur in lithium under pressure~\cite{marques2011}, may play a similar role as in the model discussed above. 

\section{Outlook}
While we saw that the BCS gap equations can be solved relatively easily for the jellium model, the model itself contains a number of approximations: (i) The Thomas-Fermi approximation was used to describe the screening of the Coulomb interaction. (ii) The frequency dependence of the interaction was treated by substituting the energy difference of the interacting electrons. As was shown by DeGennes~\cite{degennes1966} this is equivalent to treating the electron-phonon coupling in second order perturbation theory. (iii) The BCS variational wavefunction was assumed. (iv) The potential of the positively charged nuclei was replaced with a constant value. Removing some or all of these approximations would allow for more realistic modeling of the superconducting properties. This may require a radically different approach such as the holographic method~\cite{holographic} of which Jan was a great ambassador. In certain parameter ranges the static susceptibility diverges for wave vector $q_{c}$, while the interaction in the s-wave channel of the undistorted system is attractive. Within the limitations of the model used here these aspects could not be addressed more deeply, but I believe that they deserve further attention. Jan Zaanen proposed, together with Jian-Huang She, that high $T_c$ superconductivity can be obtained by using a material for which the pair susceptibility has an algebraic temperature dependence. The main finding of the present paper is that tuning the system close to a lattice instability boosts the pairing interaction. Combining these two elements, algebraic pair susceptibility and proximity to a lattice instability, looks like a promising strategy for the realization of superconductivity at room temperature. 

\section{Conclusions}
\label{section:conclusions}
We have explored DeGenne's intuitive “jellium” description of the effective interaction, which treats the Coulomb repulsion and the phonon-mediated interaction in one fell swoop. One additional element -not considered by DeGennes- was included, namely the elastic response of the charge-compensating background. When leaving out this elastic response, the interaction potential vanishes in the static limit. Nevertheless, the BCS gap equation has a solution with a non-zero $T_c$. If a positive elastic response is added (causing the sound velocity to increase), the interaction potential in the static limit is repulsive. Even in this case $T_c$ can be non-zero. If a negative elastic response is introduced (causing the sound velocity to decrease), the interaction potential in the static limit is attractive and qualitatively similar to the standard BCS scenario for pairing. This has the effect of boosting the critical temperature. 

\section{Acknowledgements}
I am grateful to Erik van Heumen, Christophe Berthod, Louk Rademaker, Frank Marsiglio and Alessio Zaccone for inspiring discussions and comments. This paper is dedicated to the memory of Jan Zaanen who has been an inexhaustible source of original perspectives.

\appendix
\section{Linear response of electrons and a compressible charge-compensating background}
\label{section:response} 
The dielectric function is defined as the linear response of the charge of the material to a test charge $\rho_{ext}$. 
The charge of the material has in the present case two contributions: electrons and the massive charge-compensating background.
We use the mean-field approximation for the dielectric function, so that it becomes the sum of two independent contributions from the electrons and the charge-compensating background
\begin{equation}
\epsilon(q,\omega)=1-{\color{black}4\pi}\chi^{(0)}_e-{\color{black}4\pi}\chi^{(0)}_{{0}}
\label{eq:epsilonA}
\end{equation}
For the electronic susceptibility we adopt the Thomas-Fermi approximation
\begin{equation}
{\color{black}4\pi}\chi^{(0)}_e=-\frac{k_{0}^2}{q^2}
\label{eq:TF}
\end{equation}
The charge-compensating background is characterized by the following properties
\begin{itemize}
\item Mass density $\rho_{m}$, charge density $\rho_{c}$ with the ratio $\rho_{c}/\rho_{m}$ fixed by the charge/mass ratio of the matter constituting the charge-compensating background, and the corresponding fluctuations of density $\delta\rho_{m,e}(\vec{r},t)$ and current $\delta\vec{j}_{m,c}(\vec{r},t)$ satisfying the continuity relation $\nabla \cdot \delta\vec{j}_{m,c}=-\partial \delta\rho_{m,c}/\partial t $.
\item the bulk modulus $B_q$, which may depend on the wave vector $q$ of the density fluctuation.
\end{itemize}
$\delta\vec{E}$ is the electric field generated by the test charge and the resulting fluctuation of the charge-compensating background respectively, {\it i.e.}
\begin{align}
\nabla \cdot \delta\vec{E} &=4\pi(\delta\rho_{ext}+\rho_{c})
\label{eq:poisson}
\end{align}
The fluctuations of mass and charge contained inside an infinitesimal volume element $\Omega$ are $M=\Omega\delta\rho_{m} = \Omega(\rho_m/\rho_c)\delta\rho_{c} $ and $Q=\Omega\delta\rho_c $. The fluctuation of the mass-flow represented by this volume element is $\Omega\delta\vec{j}_{m} =  \Omega(\rho_m/\rho_c) \delta\vec{j}_{c} $. The restoring force due to the density fluctuation is $\delta \vec{F} = -  \Omega B_q  (\vec{\nabla} \delta \rho_{m} / \rho_{m})  = -  \Omega B_q  (\vec{\nabla} \delta \rho_{c} / \rho_{c})    $.  The acceleration of the mass $M $ is given by Newton's law
\begin{equation}
M \frac{d }{dt} \delta\vec{v} = Q\delta\vec{E}  + \delta \vec{F}
\end{equation}
so that
\begin{equation}
\frac{\rho_m}{\rho_c}\frac{d }{dt} \delta\vec{j}_{c}=  \rho_{c}\delta\vec{E}  - B_q\frac{\vec{\nabla}\delta\rho_{c} }{\rho_{c}} . 
\end{equation}
Taking the divergence of both sides and using the continuity relation we obtain in the limit of small motion  
\begin{align}
&
-\frac{\partial^2}{\partial t^2}  \delta\rho_{c} = \frac{4\pi \rho_{c}^2}{\rho_m}\left(\delta\rho_{ext}  +  \delta\rho_{c}\right)    -    B_q\vec{\nabla}\cdot\vec{\nabla}\delta\rho_{c} 
\nonumber\\
\end{align}
We substitute $\delta\rho_{m,c}(\vec{r},t) = \delta\rho_{m,c} e^{i(\vec{q}\cdot\vec{r}-\omega t)} $ and obtain
\begin{align}
&
\omega^2  \delta\rho_{c}  = \omega_{{0}}^2\left(\delta\rho_{ext}  +  \delta\rho_{c}\right)  +B_q q^2 \delta\rho_{c}
\end{align}
where 
\begin{equation}
\omega_{{0}}^2=\frac{4\pi  \rho_c^2}{\rho_m}.
\end{equation}
We thus arrive at the contribution to the susceptibility of the massive charge-compensating background
\begin{equation}
{\color{black}4\pi}\chi^{(0)}_{{0}}=\frac{\delta\rho_{c}}{\delta\rho_{c}+\delta\rho_{ext}} = \frac{\omega_{{0}}^2}{\omega^2-B_q\rho_{m}^{-1}q^2}
\label{eq:chib}
\end{equation}
We adopt the following model for the bulkmodulus
\begin{equation}
B_q=\frac{{B_0}}{1+q^2s^2}
\end{equation}
where ${b_0\rho_m^{-1}}$ is usually positive, but negative values are possible. One may wonder why a $q$-dependence is introduced. It would in fact be unreasonable to assume that the elastic properties would remain the same at all length scales. The only important requirement for the present discussion is, that $B_q$ converges to zero as $1/q^2$, the details of the $q$ dependence don't matter. Substituting Eqs.~ \ref{eq:TF} and \ref{eq:chib} in Eq.~\ref{eq:epsilon} results in
\begin{equation}
\epsilon(q,\omega)=1+\frac{k_{0}^2}{q^2}+\frac{\omega_{{0}}^2}{{{B_0\rho_m^{-1}} q^2}/{(1+q^2s^2)} -\omega^2}
\end{equation}

\section*{Data and code availability}

\begin{itemize}
    \item The theoretical data generated in this study are available in Ref.~\cite{opendataLi}. These will be preserved for 10 years. All other data that support the plots within this paper and other findings of this study are available from the author upon reasonable request.
    \item The custom computer codes used to generate the results reported in this paper are available in Ref.~\cite{opendataLi}.
    \item Any additional information required to reanalyze the data reported in this paper is available from the author upon request.    
    
\end{itemize}

\section*{Declaration of interests}

The author declares no competing interests.

\end{document}